\UseRawInputEncoding
\documentclass[superscriptaddress, twocolumn, amsmath, amssymb, aps,pra, notitlepage,longbibliography]{revtex4-2}
\usepackage{graphicx}
\usepackage{dcolumn}
\usepackage{bm}
\usepackage[colorlinks=true,
            linkcolor=red,
            urlcolor= blue,
            citecolor=blue]{hyperref}
\usepackage{tikz}
\usetikzlibrary{arrows.meta, positioning, shapes.geometric}
\begin{document}

\title{Manipulation of information flow and thermodynamic performance in nonreciprocal quantum dot information engines} %

\author{Hao Feng}
\affiliation{Department of Physics, Institute for Quantum Science and Technology, Shanghai Key Laboratory of High Temperature Superconductors, International Center of Quantum and Molecular Structures, Shanghai University, Shanghai, 200444, China}
\author{Junjie Liu}
\email{jj\_liu@shu.edu.cn}
\affiliation{Department of Physics, Institute for Quantum Science and Technology, Shanghai Key Laboratory of High Temperature Superconductors, International Center of Quantum and Molecular Structures, Shanghai University, Shanghai, 200444, China}

\begin{abstract}
Quantum information engines leverage information as a thermodynamic resource to facilitate energy conversion. In the operation of such engines, the information flow between the working substance and the controller is pivotal, however, strategies for its efficient manipulation remain largely unexplored. Here, we investigate an autonomous information engine based on a double-quantum-dot setup, where a downstream dot coupled to two reservoirs acts as the working substance, and an upstream dot coupled to a single reservoir serves as the controller. By extending the second law of thermodynamics to incorporate the effects of nonreciprocal couplings between the dots and their electronic reservoirs, we develop a thermodynamic framework that allows us to demonstrate that nonreciprocity can significantly modulate the inter-dot information flow, thereby providing a robust control mechanism. We show that the influence of nonreciprocity can be equivalently understood through a mapping to an effective reciprocal system upon a reparameterization of chemical potentials and the electron–electron coupling strength. We further analyze the impact of nonreciprocity on the engine's performance and operation regime. Our findings establish nonreciprocal coupling as an effective control knob for designing and optimizing quantum dot information engines, surpassing the capabilities of conventional reciprocal configurations.
\end{abstract}

\maketitle
\date{\today}

\section{Introduction}
The rapid miniaturization of thermal devices has established quantum thermal machines as a central platform for exploring nanoscale energy conversion~\cite{Scovil.59.PRL,Seifert.12.RPP,Anders.16.CP,Benenti.17.PR,Cangemi.24.PR}. Recently, advances in quantum information thermodynamics have revealed that information is not merely a descriptor of physical states but a genuine thermodynamic resource capable of contributing directly to thermodynamic tasks~\cite{Parrondo.15.NP,Goold.16.JPA}. This interdisciplinary perspective has reshaped the development of quantum thermal machines, leading to the emergence of quantum information thermal machines~\cite{Cao.09.PRE,Sagawa.10.PRL,Toyabe.10.NP,Sagawa_2013,Mandal.13.PRL,PhysRevX.4.031015,Bauer.12.JPA,Koski.14.PRL,Still.20.PRL,Camati.16.PRL,Admon.18.PRL,Paneru.18.PRL,Saha.22.PRL,Saha.23.PRL,Fadler.23.PRL,Enrich.23.FP,Leighton.24.PRX,Aggarwal.25.PRR,Zhang.25.PRL,PhysRevResearch.5.043280,Castro.26.PRL}. These devices harness information as a thermodynamic resource, enabling energy-conversion processes that would be impossible with conventional thermal machines.

The operation of autonomous information thermal machines generally requires a bipartite architecture consisting of a controller and a controlled system that serves as the working substance. Information generated by the controller can act as a thermodynamic resource, driving the controlled system to perform nontrivial thermodynamic tasks. A central quantity in such autonomous devices is the information flow~\cite{Allahverdyan.09.JSM,PhysRevX.4.031015,Enrich.23.FP}, which quantifies the extent to which information is transferred between the coupled components. Recent studies have revealed that information flow is not merely an auxiliary diagnostic but a dynamical quantity that directly enters local entropy balances and strongly influences the efficiency and operation regime of information engines~\cite{PhysRevX.4.031015,Enrich.23.FP,PhysRevResearch.5.043280,Leighton.24.PRX,Honma.26.PRA,Maekawa.25.A}. Despite the progress, current studies have primarily focused on elucidating the thermodynamic role of information flow--understanding how it modifies conventional thermodynamic descriptions. In contrast, how to efficiently tune the magnitude of information flow and regulate it as a controllable resource remains far less explored.

In this work, we address this need by introducing nonreciprocal interactions as a control knob for manipulating information flow and the thermodynamic performance of bipartite autonomous information engines. Nonreciprocity, broadly speaking, refers to asymmetric dynamical responses in which forward and backward processes are no longer equivalent due to the breaking of time-reversal symmetry~\cite{Fruchart.26.A}. It has emerged as a powerful design principle in quantum technology~\cite{Barzanjeh.25.A}, attracting significant interest in both general theoretical descriptions and concrete experimental implementations~\cite{Metelmann.15.PRX,Hurst_2018,Huang.18.PRL,Tokura.18.NC,PhysRevLett.124.070402,PhysRevLett.132.136301,Liu_2020,Lai.20.PRA,TangJ.22.PRL,Jiao.20.PRL,Lecocq.21.PRL,Rymarz.21.PRX,WangY.23.PRXQ,Ahmadi.24.PRL,ZhangZ.22.PRA}. Nonreciprocity enables intriguing functionalities such as directional transport~\cite{Lodahl.17.N,Mandal.20.PRL,Gou.20.PRL} and quantum-limited amplification~\cite{Kamal2011}, to name just a few. Moreover, nonreciprocal interactions have been shown to strongly modify thermodynamic entropy production~\cite{Busiello.20.PRR,Pal.21.PRR,ZhangZ.23.PRR,Manzano.24.PRX}, suggesting that their influence extends naturally to thermodynamics. These developments motivate the central idea of the present study: if nonreciprocity can reshape currents and entropy production, it should also provide a promising knob for steering the transfer and utilization of information in autonomous information engines.

Specifically, we investigate an autonomous double-quantum-dot information engine in which an upstream dot acts as the controller and a downstream dot serves as the controlled system~\cite{PhysRevX.4.031015,Strasberg.13.PRL,Kutvonen.16.PRL,Leighton.24.PRX,PhysRevResearch.5.043280}. We consider nonreciprocal interactions between the dots and their respective electronic reservoirs. We develop a thermodynamic framework featuring a generalized form of the second law of thermodynamics, which explicitly shows how nonreciprocity enters the local entropy balance as an additional contribution. Our framework reduces to an existing one~\cite{PhysRevX.4.031015} when considering a conventional setup with reciprocal dot-reservoir interactions. To facilitate comparison between nonreciprocal and reciprocal setups, we further show that an appropriate reparameterization can map the nonreciprocal setup onto an effective reciprocal one, albeit with renormalized chemical potentials and electron-electron interaction strengths. This mapping yields expressions for local entropy production rates that closely resemble those for the conventional reciprocal setup with bare parameters~\cite{PhysRevX.4.031015}, thereby enabling us to understand the essential effect of nonreciprocal dot-reservoir interactions from the perspective of tuning chemical potentials and electron-electron interaction strengths. 
Using our thermodynamic framework, we show that nonreciprocity can significantly modulate the inter-dot information flow. Turning to thermodynamic performance, we further demonstrate that nonreciprocity can enhance the information-to-work conversion capability and broaden the operational window of the heat engine mode. These results identify nonreciprocal interaction as an effective control knob for designing and optimizing quantum-dot information engines.

The structure of the paper is organized as follows. In Sec.~\ref{sec:2}, we first recap the thermodynamic description of the reciprocal autonomous double-quantum-dot information engine for completeness. We then generalize the thermodynamic framework to incorporate the effects of nonreciprocal dot-reservoir interactions, and show how a reparameterization can map a nonreciprocal model onto a reciprocal one with renormalized parameters. We further introduce figures of merit to quantify the thermodynamic performance of the information engine. In Sec.~\ref{sec:3}, we present numerical results demonstrating that nonreciprocity can serve as an effective control knob to modulate the inter-dot information flow, as well as the thermodynamic performance and operating regime of the heat engine mode. Finally, we present several remarks and conclude the study in Sec.~\ref{sec:4}.

\section{Model and thermodynamic framework}\label{sec:2}
In this section, we present the theoretical details of this study, including the setup of the autonomous double-quantum-dot information engine and the associated thermodynamic expressions for the local entropy production rate, information flow, and efficiency. We pay particular attention to the distinction between reciprocal and nonreciprocal setups, as revealed by comparing thermodynamic expressions and performing a reparameterization.

\subsection{Reciprocal double-quantum-dot model revisited}
To fix notation and establish a clear baseline for our later nonreciprocal generalization, we first recap the details of an autonomous reciprocal double-quantum-dot information engine~\cite{PhysRevX.4.031015,Strasberg.13.PRL,Kutvonen.16.PRL,Leighton.24.PRX,PhysRevResearch.5.043280}. The system consists of two single-level quantum dots coupled via Coulombic repulsion of strength $U$. In this setup, a downstream dot (hereafter referred to as the $Y$ dot) acts as the working substance that extracts energy, while an upstream dot (hereafter the $X$ dot) serves as a feedback controller that manages information. To drive electron transfer, the $Y$ dot is coupled to left ($L$) and right ($R$) electron reservoirs at a uniform temperature $T_Y$. These two reservoirs have chemical potentials $\mu_L$ and $\mu_R$, respectively, creating a finite bias $\Delta\mu = \mu_L - \mu_R>0$. The state of the $Y$ dot is denoted by $y \in {0, 1}$, where $y = 1$ indicates that the energy level $\epsilon_Y$ is occupied by an electron and $y = 0$ indicates that it is empty. The $X$ dot is coupled to a single, independent electron reservoir ($D$) at temperature $T_X$ (with $T_X \leq T_Y$) and chemical potential $\mu_X$. Its state is similarly defined by $x \in {0, 1}$, with a bare energy level $\epsilon_X$. The interaction between the two dots imposes an additional electrostatic repulsion energy $U$ on the system if and only if both dots are simultaneously occupied, i.e., when the joint state is $(x, y) = (1, 1)$.

We assume that the dot–reservoir coupling strengths are weak such that a Markovian approximation is applicable. In the Markovian limit, the dynamical evolution of the double-quantum-dot system is governed by a continuous-time Markov jump process, where the probability $p(x, y)$ of the joint state $(x, y)$ evolves according to (time dependence is suppressed)
\begin{equation}
d_t p(x, y) = \sum_{x', y'} \left[ W_{x, x'}^{y, y'} p(x', y') - W_{x', x}^{y', y} p(x, y) \right].
\end{equation}
Here, $W_{x, x'}^{y, y'}$ denotes the transition rate for the transition $(x', y') \to (x, y)$. Crucially, we impose the bipartite condition~\cite{PhysRevX.4.031015}: the elements of the transition rate matrix $W$ are nonzero only for transitions with $x \neq x'$ and $y = y'$ (an $X$-only jump with rate $W_{x,x'}^y$) or $x = x'$ and $y \neq y'$ (a $Y$-only jump with rate $W_x^{y,y'}$). That is, we exclude simultaneous jumps of both dots.

We require that the transition rates satisfy local detailed balance (LDB) in the reciprocal model. For the controller ($X$ dot), the rates at which an electron enters ($W_{10}^y$) or leaves ($W_{01}^y$) the dot, conditioned on the fixed state $y$ of the working substance being fixed, are respectively given by
\begin{equation}
W_{10}^{y} = \Gamma f_{y}, \quad W_{01}^{y} = \Gamma (1 - f_{y}),
\label{eq:rates_X_reciprocal}
\end{equation}
where $\Gamma$ is the tunneling rate between the $X$ dot and its reservoir. $f_{y} = \left\{1 + \exp[(\epsilon_{X} + yU - \mu_{D})/T_{X}]\right\}^{-1}$ is an effective Fermi-Dirac distribution modulated by electron-electron coupling, which is $y$-state dependent because $U$ contributes to the energy only when both dots are occupied. Note that here we work with reciprocal dot-reservoir interaction, as manifested by the identical tunneling rate $\Gamma$ for both the $0 \to 1$ and $1 \to 0$ transitions on the $X$ dot. Similarly, for the working substance ($Y$ dot) exchanging electrons with lead $\nu \in \{L, R\}$ through reciprocal dot-reservoir couplings, the transition rates conditioned on the fixed controller's state $x$ are
\begin{equation}
W_{x}^{10,(\nu)} = \Gamma_{x}^{(\nu)} f_{x}^{(\nu)}, \quad W_{x}^{01,(\nu)} = \Gamma_{x}^{(\nu)} (1 - f_{x}^{(\nu)}),
\label{eq:rates_Y_reciprocal}
\end{equation}
where $\Gamma_{x}^{(\nu)}$ denotes the tunneling rate between the $Y$ dot and $\nu$ reservoir, and the effective Fermi-Dirac distribution $f_{x}^{(\nu)} = \{1 + \exp[(\epsilon_{Y} + xU - \mu_{\nu})/T_{Y}]\}^{-1}$ depends on the state $x$ through the electron-electron coupling. 

For such a composite system operating in a non-equilibrium steady state, the total entropy production rate (EPR), $\dot{S}_i$, is defined as~\cite{Seifert.12.RPP}
\begin{equation}\label{eq:entropy_production}
\dot{S}_{i} = \sum_{x \ge x', y \ge y'} J_{x,x'}^{y,y'} \ln \frac{W_{x,x'}^{y,y'} p(x',y')}{W_{x',x}^{y',y} p(x,y)} \ge 0,
\end{equation}
where $J_{x,x'}^{y,y'} = W_{x,x'}^{y,y'} p(x',y') - W_{x',x}^{y',y} p(x,y)$ is the steady-state probability current from state $(x', y')$ to state $(x, y)$. $\dot{S}_i$ is non-negative, thereby ensuring the compliance with the second law of thermodynamics. Since the state probabilities are constant in the steady state, the system's Shannon entropy remains unchanged. Thus, the total EPR is entirely given by the environmental EPR $\dot{S}_r$, which quantifies the heat exchanged with the reservoirs
\begin{equation}\label{eq:eq6}
\dot{S}_{r} = \sum_{x \ge x', y \ge y'} J_{x,x'}^{y,y'} \ln \frac{W_{x,x'}^{y,y'}}{W_{x',x}^{y',y}}.
\end{equation}

\begin{figure}[b!]
\centering
\begin{tikzpicture}[>=Stealth, node distance=3.5cm, thick]
    \node (00) at (0,0) [circle, draw, minimum size=1.3cm, fill=white] {$(0,0)$};
    \node (10) at (4,0) [circle, draw, minimum size=1.3cm, fill=white] {$(1,0)$};
    \node (11) at (4,4) [circle, draw, minimum size=1.3cm, fill=white] {$(1,1)$};
    \node (01) at (0,4) [circle, draw, minimum size=1.3cm, fill=white] {$(0,1)$};

    \draw[<->, blue, very thick] (00) -- (10);
    \draw[<->, blue, very thick] (11) -- (01);
    
    \draw[<->, red, very thick] (10) to[bend left=25] (11);
    \draw[<->, gray, dashed] (11) to[bend left=25] (10);

    \draw[<->, red, very thick] (01) to[bend right=25] (00);
    \draw[<->, gray, dashed] (00) to[bend right=25] (01);

    \node at (2, 2) {\Huge $\circlearrowleft$};
    \node at (2, 2) [yshift=-0.7cm] {$\mathcal{C}$};

    \node (cy0) at (-1.8, 2) {\Large $\circlearrowleft$};
    \node at (-1.8, 2) [yshift=-0.5cm] {$\mathcal{C}_Y^0$};
    \draw[gray, thin] (cy0.east) -- (0, 2);

    \node (cy1) at (5.8, 2) {\Large $\circlearrowleft$};
    \node at (5.8, 2) [yshift=-0.5cm] {$\mathcal{C}_Y^1$};
    \draw[gray, thin] (cy1.west) -- (4, 2);
\end{tikzpicture}
\caption{\label{fig:cycle_decomposition} 
Schematic representation of the cycle decomposition in the state space of double-quantum-dot system. The four joint states $(x,y)$ are connected by distinct particle exchange processes. Horizontal blue solid lines denote particle exchange with the reservoir of $X$ dot. Red solid (gray dashed) vertical curves correspond to particle exchanges with the left (right) reservoir of the $Y$ dot. The global cycle $\mathcal{C}$ is composed specifically of the transitions marked by solid lines (i.e., the processes involving the $X$ dot's reservoir and the $Y$ dot's left reservoir). Local cycles $\mathcal{C}_Y^0$ and $\mathcal{C}_Y^1$ indicate internal transitions within the $Y$ dot for a fixed state of $X$.}
\end{figure}
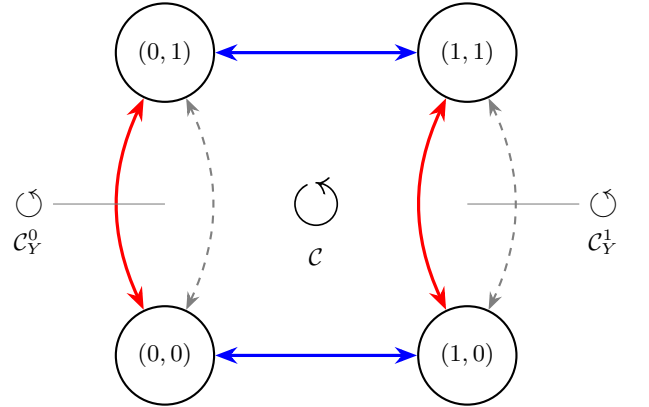

To quantify the correlations between the two quantum dots, we consider the mutual information $\mathcal{I} = \sum_{x,y} p(x,y) \ln [p(x,y)/(p(x)p(y))]$. In a bipartite setup, the time derivative of this correlation naturally splits into two directional information flows expressed in terms of conditional probabilities, $d_t \mathcal{I}=\dot{I}^{X}+\dot{I}^{Y}$, with
\begin{equation}\label{eq:inf_flow_definition}
\dot{I}^{X} = \sum_{x \ge x', y} J_{x,x'}^{y} \ln \frac{p(y|x)}{p(y|x')},~~\dot{I}^{Y} = \sum_{x, y \ge y'} J_{x}^{y,y'} \ln \frac{p(x|y)}{p(x|y')},
\end{equation}
where $\dot{I}^{X}$ denotes the rate at which the $X$ dot gains information through its own transitions, and $J_{x,x'}^y$ is the probability current from state $(x',y)$ to $(x,y)$. Similar interpretations hold for $\dot{I}^{Y}$ and $J_{x}^{y,y'}$. Under steady-state operation, the total mutual information remains constant ($d_t \mathcal{I} = 0$), leading to a continuous and balanced information flow $\dot{I} \equiv \dot{I}^{X} = -\dot{I}^{Y}$. Using information flow, one can appropriately express local entropy production rate. According to the bipartite thermodynamic formalism \cite{PhysRevX.4.031015}, the steady state local entropy production rate $\dot{S}_i^k$ for each subsystem $k \in \{X, Y\}$ is defined by the balance between local environmental EPR $\dot{S}_r^k$ and the information flow $\dot{I}$,
\begin{align}
\dot{S}_{i}^{X} &= \dot{S}_{r}^{X} - \dot{I} \ge 0,\label{eq:sxi}\\
\dot{S}_{i}^{Y} &= \dot{S}_{r}^{Y} + \dot{I} \ge 0.\label{eq:syi}
\end{align}
Here, $\dot{S}_r^X$ ($\dot{S}_r^Y$) is obtained from Eq. (\ref{eq:eq6}) by limiting to $X$-only ($Y$-only) transitions with $y=y'$ ($x'=x$).

To enable a geometric interpretation of EPR, we apply the Schnakenberg network theory~\cite{Schnakenberg.76.RMP} to decompose the steady-state probability currents into contributions from cycle fluxes associated with independent fundamental cycles, as schematically illustrated in Fig.~\ref{fig:cycle_decomposition}. For this specific bipartite device, the transition network is characterized by two distinct types of cycles~\cite{PhysRevX.4.031015}. The first type comprises the local cycles, denoted as $\mathcal{C}_{Y}^{0}$ and $\mathcal{C}_{Y}^{1}$, which represent the independent electron transport through the working substance ($Y$ dot) when the controller ($X$ dot) is frozen in the empty $(x=0)$ and occupied $(x=1)$ states, respectively. The second type is the central global cycle $\mathcal{C}$, which traces the cooperative counter-clockwise sequence of states $(0,0) \rightarrow (1,0) \rightarrow (1,1) \rightarrow (0,1) \rightarrow (0,0)$. For an oriented fundamental cycle $\mathcal{C}_{n}\in (\mathcal{C},\mathcal{C}_{Y}^{0},\mathcal{C}_{Y}^{1})$, we assign a cycle current $J(\mathcal{C}_{n})$ which quantifies the probability current circulating along the cycle $\mathcal{C}_{n}$. These cycle fluxes can be determined by the following relation~\cite{PhysRevX.4.031015}
\begin{equation}\label{eq:cycle_flux}
J^{y,y'}_{x,x'}
=
\sum_{n}
\delta^{y,y'}_{x,x'}(\mathcal{C}_{n}) J(\mathcal{C}_{n}).
\end{equation}
Here, the sum is performed over three fundamental cycles, the function $\delta^{y,y'}_{x,x'}(\mathcal{C}_{n})$ is defined as 
\begin{equation}
\delta^{y,y'}_{x,x'}(\mathcal{C}_{n}) =
\begin{cases}
+1, & \text{if } (x',y')\to(x,y) \text{ belongs to }
\mathcal{C}_{n} ,\\
-1, & \text{if } (x,y)\to(x',y') \text{ belongs to }
\mathcal{C}_{n} ,\\
0, & \text{otherwise}.
\end{cases}
\end{equation}

According to Fig.~\ref{fig:cycle_decomposition}, the global cycle flux $J(\mathcal{C})$ mediates the exchange of energy and information between the two dots. In comparison, the local cycle fluxes $J(C_Y^0)$ and $J(C_Y^1)$ represent the flow of electrons between the left and right reservoirs of the $Y$ dot. As a consequence, the steady-state information flow is exclusively carried by the global cycle $\mathcal{C}$,
\begin{equation}\label{eq:inf_flow_dec}
\dot{I} = \mathcal{J}(\mathcal{C}) \mathcal{F}^{I}(\mathcal{C}).
\end{equation}
Here, the information affinity $\mathcal{F}^{I}(\mathcal{C})$ depends purely on the steady-state conditional probabilities of the bipartite states:
\begin{equation}
\mathcal{F}^{I}(\mathcal{C}) = -\ln \frac{p(x=1|y=1)p(x=0|y=0)}{p(x=1|y=0)p(x=0|y=1)}.\label{eq:info_affinity}
\end{equation}
Substituting Eqs.~(\ref{eq:cycle_flux}) and (\ref{eq:inf_flow_dec}) into Eqs.~(\ref{eq:sxi}) and (\ref{eq:syi}), one has~\cite{PhysRevX.4.031015}
\begin{align}
\dot{S}_{i}^{X} &= \mathcal{J}(\mathcal{C}) \left[ \frac{U}{T_X} - \mathcal{F}^{I}(\mathcal{C}) \right],\label{eq:partial_S_X} \\
\dot{S}_{i}^{Y} &= -\mathcal{J}_{e} \frac{\Delta\mu}{T_Y} + \mathcal{J}(\mathcal{C}) \left[ \mathcal{F}^{I}(\mathcal{C}) - \frac{U}{T_Y} \right].\label{eq:partial_S_Y}
\end{align}
Here, $\mathcal{J}_{e}=\mathcal{J}(\mathcal{C}_Y^0) + \mathcal{J}(\mathcal{C}_Y^1)$.

Eqs.~(\ref{eq:partial_S_X}) and (\ref{eq:partial_S_Y}) encapsulate the operational condition of the device. While this setup can fundamentally operate in various thermodynamic regimes, we primarily focus on its operation as an information-powered engine. In this engine mode, the working substance ($Y$ dot) extracts useful work by pumping an electron current uphill against a chemical potential bias (i.e., $\mathcal{J}_e > 0$ for $\Delta\mu > 0$). The underlying mechanism of this information-powered work generation is transparently revealed by the local non-negative EPR. We first observe that the term $-\mathcal{J}_e\Delta\mu/T_Y$ in Eq.~(\ref{eq:partial_S_Y}) is strictly negative in the engine mode. To ensure thermodynamic consistency with $\dot{\mathcal{S}}_i^Y \ge 0$, we must require a positive information affinity $\mathcal{F}^{I}(\mathcal{C})$ to compensate for negative contributions, which thus acts as a generalized thermodynamic force driving the engine mode. Concretely, $X$ dot (the controller) acts as a sensor, dissipating thermal energy ($U/T_X$) into the cold reservoir to continuously acquire information about $Y$ dot ($\dot{I} > 0$). Subsequently, $Y$ dot consumes this incoming information flow as a thermodynamic fuel to drive an electron flow against the potential bias. 

\subsection{Nonreciprocal generalization}
\subsubsection{Model and thermodynamics}
From the preceding discussion, we see that the transition rates between states are rigidly constrained by LDB. This condition reflects the reciprocal nature of the system's coupling to equilibrium reservoirs with fixed temperatures and chemical potentials. Here, we relax this reciprocal restriction and systematically explore how continuous information flow and operation regime of heat engine mode can be actively regulated. Physically, breaking reciprocity can be realized by subjecting the system to external active feedback control or non-conservative forces that continuously inject energy to bias the transition kinetics~\cite{Fruchart.26.A}, without necessarily altering the bare energy levels of the quantum dots. We implement this strategy phenomenologically by introducing unequal tunneling rates for pairs of forward and backward transitions, thereby reflecting nonreciprocal couplings between the dots and their reservoirs.

Specifically, for the controller $X$ dot, the transition rates between the empty ($x=0$) and occupied ($x=1$) states, which are dynamically conditioned on the instantaneous state $y$ of the working substance, are generalized as
\begin{equation}\label{eq:wy_non}
    W_{10}^y = \Gamma^{y+} f_y, \quad W_{01}^y = \Gamma^{y-} (1-f_y).
\end{equation}
To quantitatively capture the asymmetry in these transition kinetics, we define the nonreciprocity ratio for $X$ dot as $\gamma^y = \Gamma^{y+}/\Gamma^{y-}$. The conventional reciprocal rates in Eq. (\ref{eq:rates_X_reciprocal}) are naturally recovered when $\gamma^y = 1$. Similarly, for the working substance ($Y$ dot) exchanging electrons with leads $\nu \in \{L, R\}$ at the background temperature $T_Y$, the transition rates conditioned on the state $x$ of $X$ dot are extended to
\begin{equation}\label{eq:wx_non}
    W_{x}^{10,(\nu)} = \Gamma_{x+}^{(\nu)} f_x^{(\nu)}, \quad W_{x}^{01,(\nu)} = \Gamma_{x-}^{(\nu)} \big(1-f_x^{(\nu)}\big).
\end{equation}
The corresponding reservoir-dependent nonreciprocity ratios for $Y$ dot are defined as $\alpha_x^{(\nu)} = \Gamma_{x+}^{(\nu)}/\Gamma_{x-}^{(\nu)}$. The reciprocal rates in Eq. (\ref{eq:rates_Y_reciprocal}) represent a special case of $\alpha_x^{(\nu)}=1$.

To show how nonreciprocal dot-reservoir couplings fundamentally modify the nonequilibrium thermodynamics of the double-quantum-dot system, we turn to EPRs. We note that the definitions in Eqs. (\ref{eq:entropy_production}), (\ref{eq:eq6}), (\ref{eq:sxi}) and (\ref{eq:syi}) remain applicable in the presence of nonreciprocal dot-reservoir couplings. Starting from these definitions and using the generalized transition rates in Eqs. (\ref{eq:wy_non}) and (\ref{eq:wx_non}), we arrive at the following generalized total steady-state EPR expressed in terms of the fundamental cycle currents
\begin{eqnarray}
    \dot{S}_i &=& \sum_{x=0,1} \mathcal{J}(C_Y^x) \left( \ln\frac{\alpha_x^{(R)}}{\alpha_x^{(L)}} - \frac{\Delta\mu}{T_Y} \right) \nonumber\\
    &&+ \mathcal{J}(C) \left( \ln\frac{\gamma^0 \alpha_1^{(L)}}{\gamma^1 \alpha_0^{(L)}} + \frac{U}{T_X} - \frac{U}{T_Y} \right) \ge 0.\label{eq:Si17}
\end{eqnarray}
For the sake of simplicity, we relegate the derivation details to Appendix~\ref{app:entropy_derivation}. It is evident from the above form that the total steady-state EPR receives contributions $\ln\frac{\alpha_x^{(R)}}{\alpha_x^{(L)}}$ and $\ln\frac{\gamma^0 \alpha_1^{(L)}}{\gamma^1 \alpha_0^{(L)}}$ that originate directly from the nonreciprocity and vanish in the reciprocal limit. For the local environmental EPRs, one obtains (see Appendix~\ref{app:entropy_derivation} for details)
\begin{eqnarray}
    \dot{S}_r^X &=& \mathcal{J}(C) \left[ \ln\frac{\gamma^0}{\gamma^1} + \frac{U}{T_X} \right],
    \label{eq:Srx_non}\\
    \dot{S}_r^Y &=& \sum_{x=0,1} \mathcal{J}(C_Y^x) \left[ \ln\frac{\alpha_x^{(R)}}{\alpha_x^{(L)}} - \frac{\Delta\mu}{T_Y} \right] \nonumber\\
    && + \mathcal{J}(C) \left[ \ln\frac{\alpha_1^{(L)}}{\alpha_0^{(L)}} - \frac{U}{T_Y} \right].
    \label{eq:Sry_non}
\end{eqnarray}
Combining these environmental contributions with the information flow according to Eqs.~(\ref{eq:sxi}) and (\ref{eq:syi}), the local EPRs in the nonreciprocal setup become (see Appendix~\ref{app:entropy_derivation} for details)
\begin{eqnarray}
    \dot{S}_i^X &=& \mathcal{J}(C) \left[ \ln\frac{\gamma^0}{\gamma^1} + \frac{U}{T_X} - \mathcal{F}^I(C) \right],\label{eq:Six18} \\
    \dot{S}_i^Y &=& \sum_{x=0,1} \mathcal{J}(C_Y^x) \left[ \ln\frac{\alpha_x^{(R)}}{\alpha_x^{(L)}} - \frac{\Delta\mu}{T_Y} \right] \nonumber \\
    && + \mathcal{J}(C) \left[ \ln\frac{\alpha_1^{(L)}}{\alpha_0^{(L)}} - \frac{U}{T_Y} + \mathcal{F}^I(C) \right].\label{eq:Siy19}
\end{eqnarray}
Eqs. (\ref{eq:partial_S_X}) and (\ref{eq:partial_S_Y}) are recovered in the reciprocal limit. These generalized expressions for the EPRs are physically revealing, as they indicate that the logarithms of the non-reciprocal ratios act on perfectly equal footing with traditional thermodynamic affinities. Just as the physical bias $\Delta\mu/T_Y$ drives the electron current, the nonreciprocal terms act as generalized, actively sustained thermodynamic forces that restructure the entropy balance. As for the information flow, we remark that the definition in Eq. (\ref{eq:inf_flow_dec}) remains applicable. However, since the EPRs change in the presence of nonreciprocity, we expect that the information flow is modified as well.

\subsubsection{Effective reciprocal systems through a reparameterization}\label{sec:2b2}
To provide further insight into the essential physics of the generalized nonreciprocal system, we present reparameterization strategies that map the nonreciprocal system onto an equivalent system with effective reciprocal couplings. For later convenience, we refer to the mapped system as the effective reciprocal system. As will become clear, this effective reciprocal system cannot be obtained from the original reciprocal one, as it features renormalized chemical potentials and electron-electron coupling strengths due to nonreciprocity. Importantly, the resulting expressions for the local EPRs after reparameterization [cf. Eqs.~(\ref{eq:sxrepar}) and (\ref{eq:syrepar})] more closely resemble those of the original reciprocal system [cf. Eqs.~(\ref{eq:partial_S_X}) and (\ref{eq:partial_S_Y})] than do their direct generalized counterparts in Eqs.~(\ref{eq:Six18}) and (\ref{eq:Siy19}). Consequently, this effective reciprocal system allows us to understand the nonreciprocal system from the perspective of tuning chemical potentials and electron-electron coupling strengths, which will facilitate our numerical simulations and interpretations.

We develop two distinct reparameterization schemes, which we dub 
$\epsilon$-independent and $\epsilon$-dependent ones, depending on whether the energy levels of the two dots are renormalized. In the main text, we focus on the $\epsilon$-independent scheme, as it yields simpler expressions for the local EPRs than the $\epsilon$-dependent scheme. The latter is relegated to Appendix~\ref{app:leadindep} for interested readers. To avoid potential misunderstanding, we stress that these reparameterization schemes should be understood as mathematical transformations between two equivalent systems; they need not possess physical reality.

In the $\epsilon$-independent scheme, the main feature is the introduction of temperature-dependent renormalized electron–electron coupling strengths arising from nonreciprocity. Specifically, for the controller ($X$ dot), we introduce the following renormalized electron-electron coupling strength $U'_X$ and the chemical potential $\mu'_X$ of its connected reservoir,
\begin{eqnarray}
    U'_X &=& U - T_X \ln \frac{\gamma^1}{\gamma^0},\label{eq:UX_eff} \\
    \mu'_X &=& \mu_X + T_X \ln \gamma^0.\label{eq:muX_eff}
\end{eqnarray}
From the above expressions, it is evident that the renormalized parameters acquire a temperature dependence due to nonreciprocity as characterized by $\gamma^{0/1}\neq 1$, and recover their bare values in the reciprocal limit of $\gamma^{0/1}= 1$. Under this reparameterization, the effective Fermi-Dirac distribution for $X$ dot is constructed as $f'_y = \{1 + \exp[(\epsilon_X + yU'_X - \mu'_X)/T_X]\}^{-1}$. We then find that the generalized transition rates of the controller in Eq. (\ref{eq:wy_non}) can be compactly rewritten in an equivalent form that satisfies an effective LDB,
\begin{equation}\label{eq:e22}
    W_{10}^y = \Gamma^{y'} f'_y, \quad W_{01}^y = \Gamma^{y'} (1 - f'_y).
\end{equation}
Here, the effective reciprocal tunneling rate is $\Gamma^{y'} = \Gamma^{y-} + (\Gamma^{y+} - \Gamma^{y-})f_y$. Although the above forms resemble those of the original reciprocal model given in Eq. (\ref{eq:rates_X_reciprocal}), we remark that they are fundamentally distinct, as Eq. (\ref{eq:e22}) involves temperature-dependent renormalized parameters.

A similar reparameterization can be constructed for the working substance ($Y$ dot). Noting that $Y$ dot couples to two reservoirs (indexed by $\nu \in \{L, R\}$), we introduce the following temperature-dependent renormalized parameters,
\begin{eqnarray}
    \mu'_L = \mu_L + T_Y \ln \alpha_0^{(L)}, & \quad U'_L = U - T_Y \ln \frac{\alpha_1^{(L)}}{\alpha_0^{(L)}},\label{eq:Yeff_L} \\
    \mu'_R = \mu_R + T_Y \ln \alpha_0^{(R)}, & \quad U'_R = U - T_Y \ln \frac{\alpha_1^{(R)}}{\alpha_0^{(R)}}\label{eq:Yeff_R}.
\end{eqnarray}
Here, the renormalized chemical potentials $\mu'_\nu$ transform similarly to $\mu'_X$ but with different temperature and nonreciprocity ratios. What is different is that we need reservoir-specified renormalized electron-electron coupling strengths $U'_{\nu}$ to account for transitions involving different reservoirs. In the reciprocal limit of $\alpha_x^{(\nu)}=1$ ($x=0,1$), these renormalized parameters reduce to their bare counterparts. The corresponding renormalized Fermi-Dirac distributions are given by $f'^{(\nu)}_x = \{1 + \exp[(\epsilon_Y + xU'_\nu - \mu'_\nu)/T_Y]\}^{-1}$. With these renormalized parameters, we can show that the generalized transition rates in Eq. (\ref{eq:wx_non}) can be equivalently transformed into
\begin{equation}
W^{10,\nu}_x = \Gamma'^{(\nu)}_x f'^{(\nu)}_x, \quad W^{01,\nu}_x = \Gamma'^{(\nu)}_x(1- f'^{(\nu)}_x).
\end{equation}
Here, $\Gamma'^{(\nu)}_x = W^{10,\nu}_x + W^{01,\nu}_x = \Gamma^{(\nu)}_{x-} + (\Gamma^{(\nu)}_{x+} - \Gamma^{(\nu)}_{x-}) f_x^{(\nu)}$ denote effective reciprocal tunneling rates for $Y$ dot. The distinction from Eq. (\ref{eq:rates_Y_reciprocal}) is also transparent.

Substituting these renormalized parameters back into Eqs.~(\ref{eq:Six18}) and (\ref{eq:Siy19}) for the local EPRs derived in the previous subsection yields the following equivalent forms,
\begin{eqnarray}
    \dot{S}_i^X &=& \mathcal{J}(C) \left[ \frac{U'_X}{T_X} - \mathcal{F}^I(C) \right],\label{eq:sxrepar}\\
    \dot{S}_i^Y &=& -\mathcal{J}_e\frac{\Delta \mu'}{T_Y}+\sum_{x=0,1} \mathcal{J}(C_Y^x) \frac{x\Delta U'}{T_Y}\nonumber\\
    &&+ \mathcal{J}(C) \left( \mathcal{F}^I(C) - \frac{U'_L}{T_Y} \right),\label{eq:syrepar}
\end{eqnarray}
where we have denoted the differences $\Delta U' \equiv U'_L - U'_R$ and $\Delta \mu' \equiv \mu'_L - \mu'_R$. The cycle fluxes and information affinity remain unaltered. We observe that Eq.~(\ref{eq:sxrepar}) closely resembles Eq.~(\ref{eq:partial_S_X}) of the original reciprocal model, which allows us to understand the impact of nonreciprocity on the local EPR $\dot{S}_i^X$ solely from the renormalization of the electron-electron coupling strength $U'_X$. Comparing Eq.~(\ref{eq:sxrepar}) with Eq.~(\ref{eq:partial_S_Y}) of the original reciprocal model highlights that the impact of nonreciprocity on local EPR $\dot{S}_i^Y$ can be captured by the renormalized chemical potential bias and the reservoir-specified renormalized electron-electron coupling strengths.

While Eqs.~(\ref{eq:sxrepar}) and (\ref{eq:syrepar}) demonstrates that the impact of nonreciprocity can be equivalently understood in terms of temperature-dependent renormalized chemical potentials and electron-electron coupling strengths, these renormalized parameters play different phenomenological roles. A renormalized chemical potential bias $\Delta\mu'$ acts as a static background force; any thermodynamic effects driven purely by renormalized $\mu_{L,R,D}^{\prime}$ can be trivially realized in the conventional reciprocal system simply by adjusting the values of physical chemical potentials ($\mu_L$, $\mu_R$, and $\mu_X$) of the reservoirs. Consequently, variations in chemical potentials do not represent behaviors strictly unique to nonreciprocity. In contrast, the definitive physical signature of broken reciprocity lies in the decoupling of the electron-electron coupling strength. In a conventional reciprocal system, the electron-electron coupling strength $U$ is a single, globally fixed parameter experienced by all subsystems and transitions. Nonreciprocal interactions effectively fracture this property, enforcing the introduction of renormalized electron-electron coupling strengths that take on distinct, mutually unequal values depending on the specific reservoir involved ($U_X^{\prime} \neq U_L^{\prime} \neq U_R^{\prime}$). This emergence of multiple, independent effective electron-electron coupling strengths from a single physical coupling cannot be replicated by tuning the static biases of any reciprocal counterpart. 

To rigorously isolate and systematically explore the novel physical behaviors generated exclusively by nonreciprocity, it is logically justified to treat the parameters that possess direct reciprocal equivalents (such as $\mu^{\prime}$ and $\Gamma^{\prime}$) as a fixed baseline. The essential nonreciprocal phenomenology can then be fully captured and investigated by focusing solely on the parameter subspace spanned by the decoupled renormalized electron-electron coupling strengths $U_X^{\prime}$, $U_L^{\prime}$, and $U_R^{\prime}$. This is precisely the strategy we adopt in the numerical simulations presented in Sec. \ref{sec:3}.

\subsection{Quantifying thermodynamic performance of engine mode}\label{sec:2c}
In this subsection, we introduce figures of merit to quantify the thermodynamic performance of the nonreciprocal bipartite system operating as an information-powered heat engine. We motivate our definitions based on two considerations: (i) They should subsume the definitions used in the reciprocal system when nonreciprocity is turned off, thereby enabling a direct comparison between reciprocal and nonreciprocal settings. (ii) In defining the thermodynamic efficiency, we adopt the perspective that the efficiency of any engine (including information-assisted ones) operating with hot and cold heat baths should remain fundamentally constrained by the Carnot efficiency, provided all energy contributions are appropriately accounted for. This perspective is consistent with the development of information thermodynamics~\cite{Parrondo.15.NP}, which has shown that systems with information contributions remain constrained by the second law of thermodynamics.

In the reciprocal setup, the heat currents can be identified from the local environmental EPRs through the relations $\dot{Q}_k=-T_k\dot{S}_r^k$ for $k=X,Y$~\cite{PhysRevX.4.031015}. In the nonreciprocal setups, we still adopt the local environmental EPRs to define effective heat currents as
\begin{eqnarray}
    \dot{\mathcal{Q}}_{X} &\equiv& -T_X\dot{S}_r^X \nonumber\\
    &=& -\mathcal{J}(C)\left(U+T_X\ln\frac{\gamma^0}{\gamma^1}\right)\nonumber\\
    &=& -\mathcal{J}(C)U'_X,
    \label{eq:Qx_eff}
\end{eqnarray}
\begin{eqnarray}
    \dot{\mathcal{Q}}_{Y} &\equiv& -T_Y\dot{S}_r^Y \nonumber\\
    &=& \sum_{x=0,1}\mathcal{J}(C_Y^x)
    \left(\Delta\mu-T_Y\ln\frac{\alpha_x^{(R)}}{\alpha_x^{(L)}}\right) \nonumber\\
    &&+\mathcal{J}(C)\left(U-T_Y\ln\frac{\alpha_1^{(L)}}{\alpha_0^{(L)}}\right) \nonumber\\
    &=& \sum_{x=0,1}\mathcal{J}(C_Y^x)(\Delta\mu'-x\Delta U') \nonumber\\
    &&+\mathcal{J}(C)U'_L,
    \label{eq:Qy_eff}
\end{eqnarray}
where Eqs.~(\ref{eq:Srx_non}) and (\ref{eq:Sry_non}) have been used to obtain the second equalities of both equations, and the $\epsilon$-independent reparameterization scheme [cf. Eqs.~(\ref{eq:UX_eff}), \eqref{eq:muX_eff}, (\ref{eq:Yeff_L}), and (\ref{eq:Yeff_R})] introduced previously has been used to obtain the third equalities of both equations. In the reciprocal limit, these effective heat currents reduce to $\dot{Q}_{X,Y}$.

The second equalities of both Eqs. (\ref{eq:Qx_eff}) and (\ref{eq:Qy_eff}) clearly highlight the contributions to effective heat currents: The terms proportional to $U$ and $\Delta\mu$ reproduce the standard heat currents expected from the conventional reciprocal information engine, while the logarithmic terms arise from the nonreciprocal ratios and represent the additional energetic dissipation associated with maintaining asymmetric transition kinetics. From the third equalities of Eqs. (\ref{eq:Qx_eff}) and (\ref{eq:Qy_eff}) where we re-express the quantities in terms of reparameterized variables introduced in the previous subsection, we see that the effective heat currents restore forms similar to their counterparts in the reciprocal system, since the reparameterization scheme allows the system to satisfy an effective LDB. To fix the sign convention, $\dot{\mathcal{Q}}_Y > 0$ indicates a net heat absorption by the $Y$ dot from the hot reservoir, while $\dot{\mathcal{Q}}_X < 0$ signifies a net heat dissipation by the $X$ dot to the cold reservoir.

Substituting the effective heat currents into the local second law of thermodynamics Eqs. (\ref{eq:sxi}) and (\ref{eq:syi}) yields the fundamental constraints imposed by the continuous information flow $\dot{I}$:
\begin{eqnarray}\label{eq:322}
    -\dot{\mathcal{Q}}_X \ge T_X \dot{I},\nonumber \\
    \dot{I} \ge \frac{\dot{\mathcal{Q}}_Y}{T_Y}.
\end{eqnarray}
These relations emphasize that information serves as a thermodynamic currency: its generation in the $X$ dot requires a minimum energy cost, while its consumption in the $Y$ dot provides the necessary fuel to pump electrons against a bias.

In the steady state of such autonomous engines, the first law of thermodynamics imposes the following energetic compensation condition
\begin{equation}
    \dot{W} = \dot{\mathcal{Q}}_X + \dot{\mathcal{Q}}_Y.
\end{equation}
In the reciprocal limit, $\dot{W}$ reduces to the conventional power. In the presence of nonreciprocal dot-reservoir couplings, we regard $\dot{W}$ as an effective power that incorporates the energetic contributions from maintaining the asymmetric transition kinetics by a hidden external agent not included in the modeling. $\dot{W}>0$ marks an output power. Interestingly, we can utilize the inequalities in Eq.~(\ref{eq:322}) to obtain an upper bound for $\dot{W}$, 
\begin{equation}
    \dot{W} \le \dot{\mathcal{Q}}_Y \left( 1 - \frac{T_X}{T_Y} \right).
\end{equation}
This inequality motivates us to define the thermodynamic efficiency $\eta$ as
\begin{equation}
  \eta \equiv \dot{W} / \dot{\mathcal{Q}}_Y,  
\end{equation}
since it immediately follows that
\begin{equation}
    \eta \le 1 - \frac{T_X}{T_Y} \equiv \eta_C.\label{eq:carnot_bound}
\end{equation}
Here, $\eta_C$ is the Carnot efficiency. This definition ensures that while non-reciprocity alters the local entropy and energy balance, the efficiency remains strictly bounded by the Carnot efficiency--a perspective that we adopt. We remark that this efficiency definition applies to the global system consisting of the two dots as a whole, since it involves both heat exchanges between the two dots and their reservoirs.

To further quantify the continuous information flow's role as a thermodynamic resource, we connect our performance metrics with the bipartite information thermodynamics framework introduced in~\cite{PhysRevX.4.031015}. We adopt their definition of the information thermodynamic efficiencies, which characterize the effectiveness of information generation in the controller and its utilization in the working substance. For the controller (dot $X$), the efficiency of encoding information is bounded by the heat dissipated to the cold reservoir~\cite{PhysRevX.4.031015}:
\begin{equation}
    \epsilon^{X} = \frac{\dot{I}}{\dot{S}_{r}^{X}} = \frac{\dot{I}}{-\dot{\mathcal{Q}}_{X}/T_{X}} \le 1.
\end{equation}
Conversely, for the working substance ($Y$ dot), the efficiency of exploiting this continuous information to absorb heat from the hot reservoir is~\cite{PhysRevX.4.031015}:
\begin{equation}
    \epsilon^{Y} = \frac{|\dot{S}_{r}^{Y}|}{\dot{I}} = \frac{\dot{\mathcal{Q}}_{Y}/T_{Y}}{\dot{I}} \le 1.
\end{equation}

\begin{table*}[t]
\caption{\label{tab:numerical_parameters} Parameters and their default values used in Sec.~\ref{sec:3}. $\nu=L,R$, $y=0,1$, $x=0,1$}
\renewcommand{\arraystretch}{1.12}
\begin{tabular*}{\textwidth}{@{\extracolsep{\fill}}lll}
\hline\hline
\parbox[t]{0.19\textwidth}{\raggedright Meaning} &
\parbox[t]{0.30\textwidth}{\raggedright Parameters} &
\parbox[t]{0.43\textwidth}{\raggedright Default values} \\
\hline
\parbox[t]{0.19\textwidth}{\raggedright Reservoirs' and dots' parameters} &
\parbox[t]{0.30\textwidth}{\raggedright $\mu_X$, $\mu_{\nu}$, $T_X$, $T_Y$, $\epsilon_X$, $\epsilon_Y$} &
\parbox[t]{0.43\textwidth}{\raggedright $\mu_X=1-U/2$, $\mu_L=1.1$, $\mu_R=0.9$, $T_X=0.1$, $T_Y=1$, $\epsilon_X=\epsilon_Y=1$} \\[1.2em]
\parbox[t]{0.19\textwidth}{\raggedright $X$ dot's reciprocal tunneling rates} &
\parbox[t]{0.30\textwidth}{\raggedright $\Gamma^y$} &
\parbox[t]{0.43\textwidth}{\raggedright $\Gamma^0=\Gamma^1=100$} \\[1.2em]
\parbox[t]{0.19\textwidth}{\raggedright $Y$ dot's reciprocal tunneling rates} &
\parbox[t]{0.30\textwidth}{\raggedright $\Gamma_x^{(\nu)}$} &
\parbox[t]{0.43\textwidth}{\raggedright $\Gamma_0^{(L)}=\Gamma_1^{(R)}=1.5$, $\Gamma_1^{(L)}=\Gamma_0^{(R)}=0.5$} \\[1.2em]
\parbox[t]{0.19\textwidth}{\raggedright Nonreciprocal tunneling rates} &
\parbox[t]{0.30\textwidth}{\raggedright $\Gamma^{y\pm},\Gamma_{x\pm}^{(\nu)}$} &
\parbox[t]{0.43\textwidth}{\raggedright Varying one rate at a time while fixing others to reciprocal values.} \\[1.2em]
\parbox[t]{0.19\textwidth}{\raggedright Nonreciprocity ratios} &
\parbox[t]{0.30\textwidth}{\raggedright $\gamma^y=\Gamma^{y+}/\Gamma^{y-}$, \quad $\alpha_x^{(\nu)}=\Gamma_{x+}^{(\nu)}/\Gamma_{x-}^{(\nu)}$} &
\parbox[t]{0.43\textwidth}{\raggedright Varying numerators while fixing denominators to reciprocal values.} \\[1.2em]
\parbox[t]{0.19\textwidth}{\raggedright Interaction energy ratio} &
\parbox[t]{0.30\textwidth}{\raggedright $\lambda_{\nu}=U_{\nu}'/U$} &
\parbox[t]{0.43\textwidth}{\raggedright Varying $\lambda_{L}$ with $\lambda_{R}=1$ and vice versa.} \\[0.6em]
\hline\hline
\end{tabular*}
\end{table*}
Using these microscopic information efficiencies, we can express the macroscopic heat flows directly in terms of the information flow: $\dot{\mathcal{Q}}_{X} = -T_{X}\dot{I}/\epsilon^{X}$ and $\dot{\mathcal{Q}}_{Y} = T_{Y}\epsilon^{Y}\dot{I}$. Substituting these into the global first law, $\dot{W} = \dot{\mathcal{\mathcal{Q}}}_{X} + \dot{Q}_{Y}$, we can evaluate the performance of the engine per unit of information processed. We define this information-to-work conversion efficacy as $\xi \equiv \dot{W}/\dot{I}$, which yields:
\begin{equation}
    \xi = T_{Y}\epsilon^{Y} - \frac{T_{X}}{\epsilon^{X}}.
\end{equation}
Because the information generation and consumption are strictly bounded ($\epsilon^{X} \le 1, \epsilon^{Y} \le 1$), this conversion efficacy is universally constrained by the temperature difference between the two baths:
\begin{equation}
    \xi \le T_{Y} - T_{X}.\label{eq:xi_bound}
\end{equation}
This fundamental inequality rigorously demonstrates that the maximum rate at which the non-reciprocal engine can convert abstract information into work is intrinsically limited by the thermal gradient.

Furthermore, substituting the information-dependent heat flows into the macroscopic thermodynamic efficiency $\eta = \dot{W}/\dot{\mathcal{Q}}_{Y} = 1 + \dot{\mathcal{Q}}_{X}/\dot{\mathcal{Q}}_{Y}$ provides a direct, exact relationship between the global engine efficiency and the local information efficiencies:
\begin{equation}
    \eta = 1 - \frac{T_{X}}{T_{Y}\epsilon^{X}\epsilon^{Y}}.
\end{equation}
Since $\epsilon^{X}\epsilon^{Y} \le 1$, this exact relation gracefully recovers the Carnot bound, $\eta \le 1 - T_{X}/T_{Y} = \eta_{C}$. Crucially, it reveals that the non-reciprocal information engine can only approach the Carnot limit when both the measurement and the feedback processes operate at their respective reversible limits (i.e., $\epsilon^{X} \to 1$ and $\epsilon^{Y} \to 1$). Any internal dissipation during the continuous information exchange inherently limits the global performance.

\section{Numerical Results}\label{sec:3}
We now turn to the numerical analysis of the nonreciprocal quantum-dot information engine. The purpose of this section is to clarify how the nonreciprocal transition rates, and equivalently the renormalized electron-electron coupling strengths introduced by the reparameterization scheme, modify the steady-state information transfer and the resulting thermodynamic performance.

Before presenting the numerical results, we clarify how we vary parameters for demonstration purposes. One may either vary the original nonreciprocal tunneling rates in Eqs.~(\ref{eq:wy_non}) and (\ref{eq:wx_non}), or use the reparameterized description in terms of the renormalized electron-electron coupling strengths $U_X'$, $U_L'$, and $U_R'$. These two descriptions are related by the parameterization in Sec.~\ref{sec:2b2} and therefore represent the same underlying physics rather than independent mechanisms. In the following, we use the former for evaluating the information flow and local EPRs, and the latter for assessing thermodynamic performance, according to which representation displays the nonreciprocal effect more clearly. In Table~\ref{tab:numerical_parameters}, we list all the involved parameters, their default values, and detail how we vary them. Unless otherwise stated, the reservoirs' bare  chemical potentials and temperatures, dots' bare energies, and reciprocal tunneling rates are kept fixed at the values listed in the Table.

\begin{figure*}[t]
    \centering
    \includegraphics[width=0.82\textwidth]{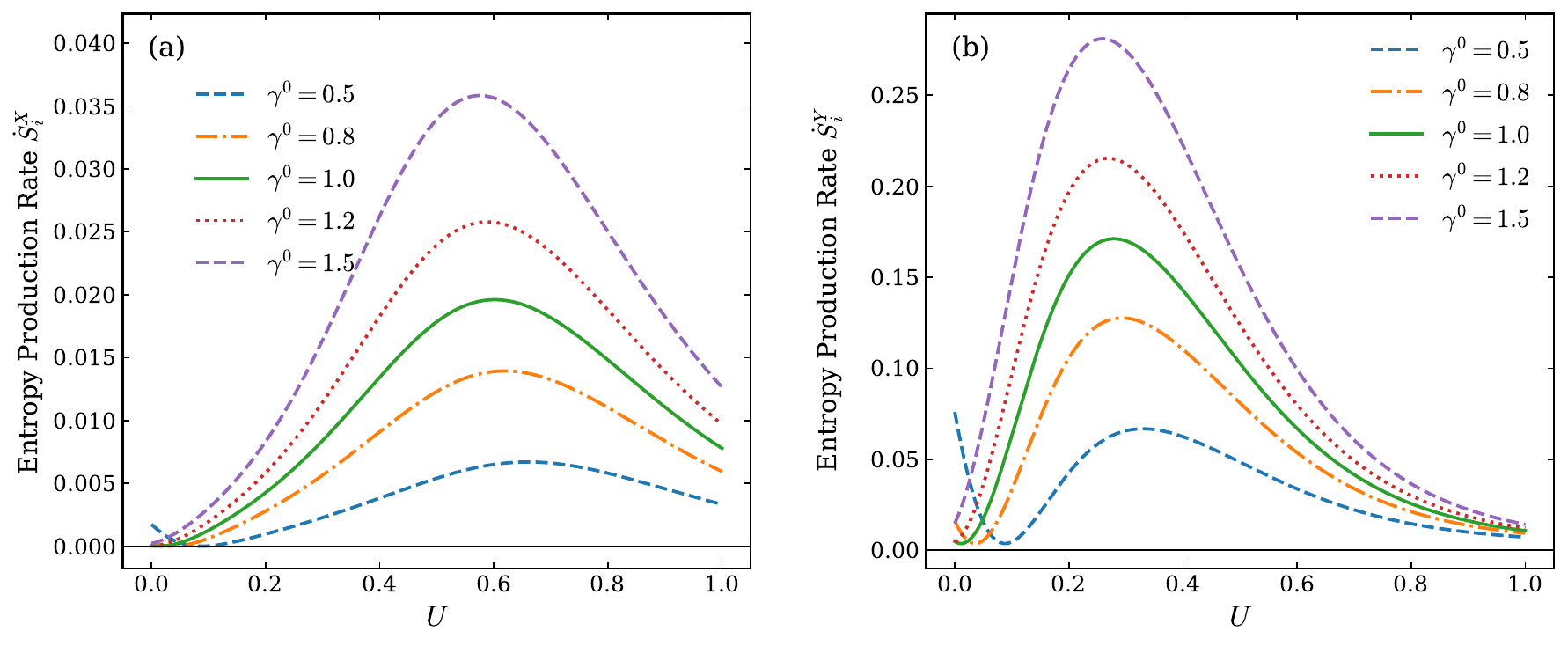}
    \caption{
    Local entropy production rates as functions of the bare electron-electron coupling strength $U$ with varying nonreciprocal ratio $\gamma^0=\Gamma^{0+}/\Gamma^{0-}$ (see the variation convention in Table~\ref{tab:numerical_parameters}).
    (a) Entropy production rate $\dot{S}_i^X$ of the controller.
    (b) Entropy production rate $\dot{S}_i^Y$ of the working substance. All other tunneling rates are fixed to their reciprocal values given in Table~\ref{tab:numerical_parameters}. 
    }
    \label{fig:epr_alpha}
\end{figure*}

\subsection{Thermodynamic consistency}
\label{sec:3B}
We first verify that the nonreciprocal modification of the transition rates is thermodynamically consistent in the sense that the local EPRs in the nonreciprocal system remain non-negative, as required by the second law of thermodynamics. This check is crucial because the local EPRs include the additional logarithmic terms associated with the nonreciprocal coupling, as shown in Sec.~\ref{sec:2}. Based on the results in Sec.~\ref{sec:2}, we remark that the local EPRs can be computed using either the expressions in Eqs.~\eqref{eq:Six18} and \eqref{eq:Siy19} or, equivalently, the expressions in Eqs.~\eqref{eq:sxrepar} and \eqref{eq:syrepar}, where the nonreciprocal contribution is absorbed into the effective electron–electron coupling strengths and effective chemical potentials. We have checked numerically that these two evaluations coincide for the same set of parameters. This agreement confirms that the two descriptions are related by a change of parametrization rather than by a change in the underlying dynamics.

Figure~\ref{fig:epr_alpha} displays a set of numerical results for the local EPRs $\dot{S}_i^X$ and $\dot{S}_i^Y$ in which $\gamma^0$ is varied, meaning that we use Eqs.~\eqref{eq:Six18} and \eqref{eq:Siy19} for the calculation. From the figure, we observe that both $\dot{S}_i^X$ and $\dot{S}_i^Y$ remain non-negative over the entire range of $U$ considered, thereby demonstrating the thermodynamic consistency of the nonreciprocal system. The controller's local EPR $\dot{S}_i^X$ is small at weak couplings, increases as the electron-electron interaction begins to generate appreciable correlations, and then decreases after the relevant cycle current is suppressed at stronger coupling. The working substance's local EPR $\dot{S}_i^Y$ shows a similar trend but with a larger magnitude, reflecting the additional dissipative contribution associated with electron transport against the chemical potential bias.


Generally speaking, the additional contributions from nonreciprocal couplings in Eqs.~\eqref{eq:Six18} and \eqref{eq:Siy19} are not optional corrections. They represent the entropy production required to maintain the asymmetric transition kinetics. Omitting them would incorrectly treat the nonreciprocal dynamics as if they were generated by ordinary reciprocal couplings and could lead to an apparent violation of the local second law of thermodynamics. Including these terms restores the correct thermodynamic accounting and ensures
\begin{equation}
    \dot{S}_i^X \ge 0, \qquad \dot{S}_i^Y \ge 0,
\end{equation}
for the nonreciprocal system.

\begin{figure*}[t]
    \centering
    \includegraphics[width=1\textwidth]{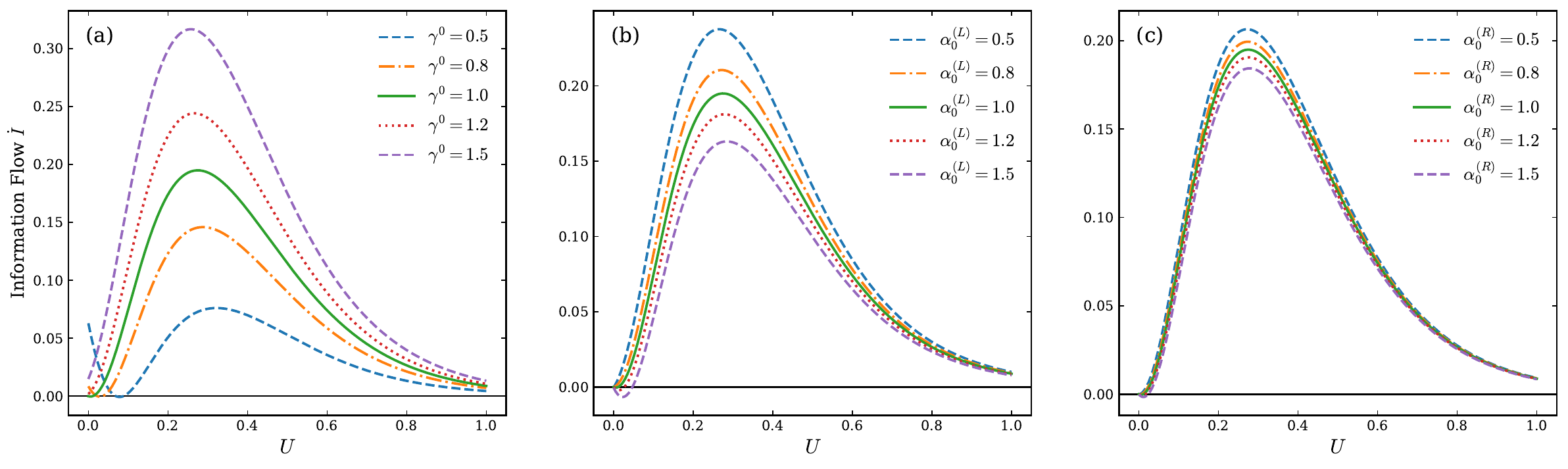}
    \caption{
    Steady-state information flow $\dot{I}$ as a function of the bare electron-electron coupling strength $U$. (a) Varying $\gamma^0$, obtained by varying $\Gamma^{0+}$, 
    (b) Varying $\alpha_{0}^{(L)}$, obtained by varying $\Gamma_{0+}^{(L)}$, and
    (c) Varying $\alpha_{0}^{(R)}$, obtained by varying $\Gamma_{0+}^{(R)}$. All other tunneling rates are fixed to their reciprocal values given in Table~\ref{tab:numerical_parameters}.}
    \label{fig:info_alpha}
\end{figure*}

\subsection{Behavior of information flow}
\label{sec:3A}
We then turn to the information flow, since it is the central quantity that connects the microscopic bipartite dynamics with the local entropy balances of the two dots. For each set of parameters, we obtain the stationary distribution by solving the master equation in the long-time limit and then evaluate the steady-state information flow according to Eq.~\eqref{eq:inf_flow_dec}, using the global-cycle current in Eq.~\eqref{eq:cycle_flux} and the information affinity in Eq.~\eqref{eq:info_affinity}. 

Figure~\ref{fig:info_alpha} shows a set of representative results for the information current as a function of the bare electron-electron coupling strength with varying nonreciprocity ratios, chosen to illustrate that the continuous information flow can be substantially modulated by nonreciprocal dot-reservoir couplings. Several general features can be extracted from Fig.~\ref{fig:info_alpha}. First, for all representative scans shown here, $\dot{I}$ exhibits a single peak when varying the electron-electron coupling strength $U$. This non-monotonic behavior reflects the competition between correlation generation and kinetic suppression. In the weak-coupling regime, the state of one dot only weakly affects the transition statistics of the other. The conditional probabilities entering $\mathcal{F}^{I}(\mathcal{C})$ are therefore only weakly separated, and the information affinity remains small. As $U$ increases, the electron-electron coupling makes the transition rates more strongly conditioned on the state of the other dot, enhancing the correlation generated along the global cycle and thereby increasing $\dot{I}$. For sufficiently large $U$, however, the joint occupation state becomes energetically costly, and the global cycle current is suppressed because part of the cycle becomes dynamically unfavorable. The information flow then decreases. The maximum of $\dot{I}$ consequently occurs at an intermediate interaction strength, where the system achieves the best compromise between strong inter-dot correlation and sufficient dynamical accessibility of the information-carrying cycle.

Second, changing a nonreciprocal tunneling rate can alter both the magnitude and the position of the maximum of the information flow. This demonstrates that nonreciprocity provides a practical kinetic handle for controlling continuous information flow. Comparing Fig.~~\ref{fig:info_alpha} (a) with (b) and (c), we observe two differences between tuning $\gamma^0$ and tuning $\alpha_0^{(L),(R)}$: (i) Tuning $\gamma^0$, which characterizes the nonreciprocal coupling between the controller dot and its reservoir, has the most significant impact on modulating the magnitude of the information flow, consistent with the indispensable role of the controller dot in generating information. (ii) Tuning $\gamma^0$ and tuning $\alpha_0^{(L),(R)}$ have opposite effects on the magnitude of information flow. For instance, decreasing $\gamma^0$ suppresses the information flow, whereas decreasing $\alpha_0^{(L),(R)}$ instead increases it. Hence, Fig.~\ref{fig:info_alpha} clearly demonstrates that even when only one of the twelve tunneling rates is changed, the resulting nonreciprocity can visibly reshape the information flow. 


\begin{figure*}[t]
    \centering
    \includegraphics[width=0.85\textwidth]{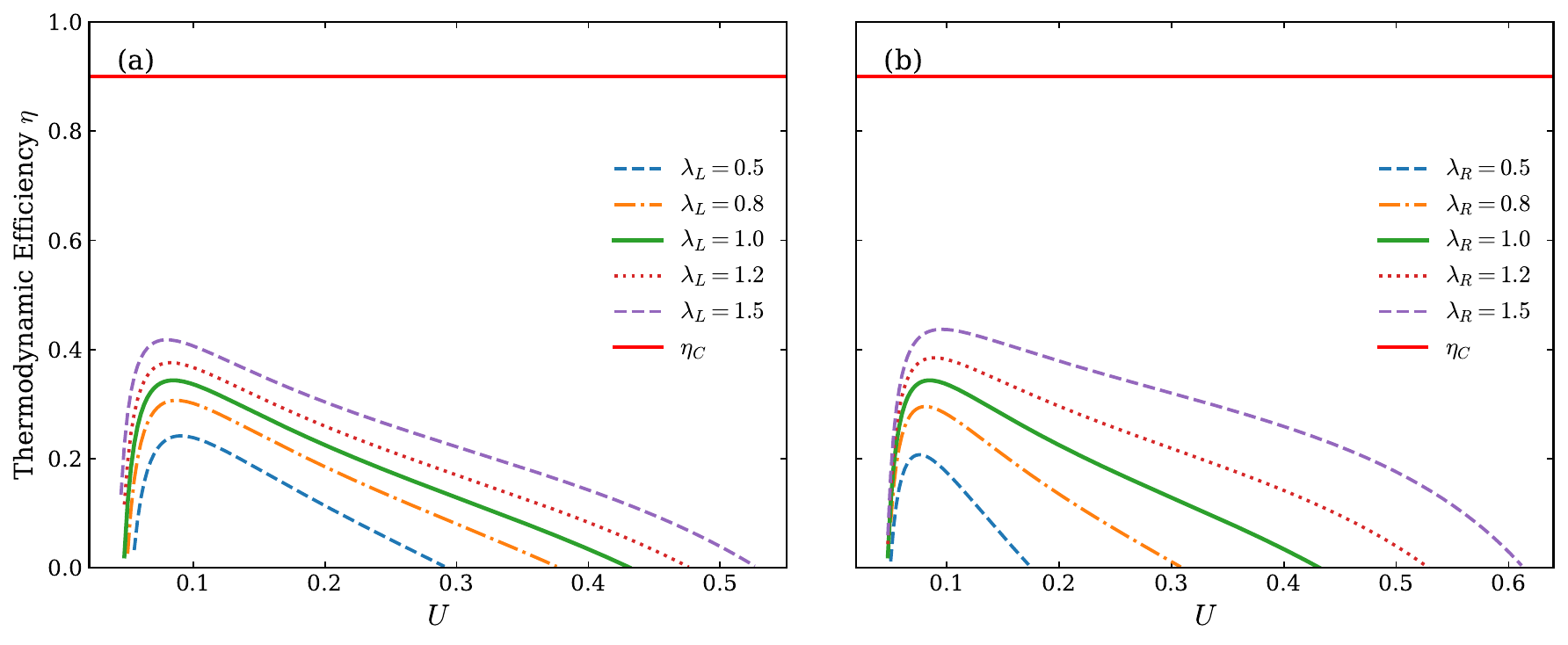}
    \caption{
    Thermodynamic efficiency $\eta$ under modulation of the renormalized electron-electron coupling strengths.
    (a) Varying $U_L'=\lambda_LU$ with $\lambda_R=1$.
    (b) Varying $U_R'=\lambda_RU$ with $\lambda_L=1$.
    In both plots, the reciprocal reference (green solid curve) corresponds to $\lambda_L=\lambda_R=1$, and the horizontal red solid line denotes the Carnot efficiency $\eta_C$. Curves are shown only in the heat-engine operation regime. All other parameters are fixed to their default values given in Table~\ref{tab:numerical_parameters}. 
    }
    \label{fig:eff_Uprime}
\end{figure*}

\subsection{Behavior of thermodynamic efficiency}
\label{sec:3C}
We next consider the thermodynamic efficiency of the double-quantum-dot system operating as a heat engine. We use the efficiency defined in Sec.~\ref{sec:2}, $\eta=\dot{W}/\dot{\mathcal{Q}}_Y$, where $\dot{W}=\dot{\mathcal{Q}}_X+\dot{\mathcal{Q}}_Y$ and effective heat currents $\dot{\mathcal{Q}}_{X,Y}$ can be calculated using Eqs. (\ref{eq:Qx_eff}) and (\ref{eq:Qy_eff}). 
To examine the behavior of the efficiency, we choose to vary the renormalized electron–electron coupling strengths $U'_{L,R}$ defined in Sec.~\ref{sec:2b2}, as this provides a clear view of how nonreciprocal couplings reshape both the magnitude of the efficiency and the operating regime of the heat-engine mode. 


Fig.~\ref{fig:eff_Uprime} displays a set of numerical results for $\eta$ with varying ratios $\lambda_L=U'_L/U$ and $\lambda_R=U'_R/U$. For a fixed ratio $\lambda_L$ (or $\lambda_R$) and varying $U$ as considered in the plot of Fig.~\ref{fig:eff_Uprime}, we adjust the nonreciprocity ratios $\alpha_x^{(\nu)}$ to vary $U'_{L}$ (or $U'_R$) according to definitions in Eqs. (\ref{eq:Yeff_L}) and (\ref{eq:Yeff_R}). We remark that only regimes where $\eta$ remains positive are shown. From the figure, we clearly observe that the efficiency remains strictly below the Carnot bound in Eq.~\eqref{eq:carnot_bound}. This confirms our expectation that the efficiency of nonreciprocal systems remains constrained by the thermodynamics. More interestingly, we see that increasing the ratios $\gamma_{L,R}>1$ can substantially broaden the operation regime of the heat-engine mode together with an enhanced efficiency. This provides a clear guideline for optimizing the double-quantum-dot heat engine using nonreciprocal couplings.

\begin{figure*}[t]
    \centering
    \includegraphics[width=0.87\textwidth]{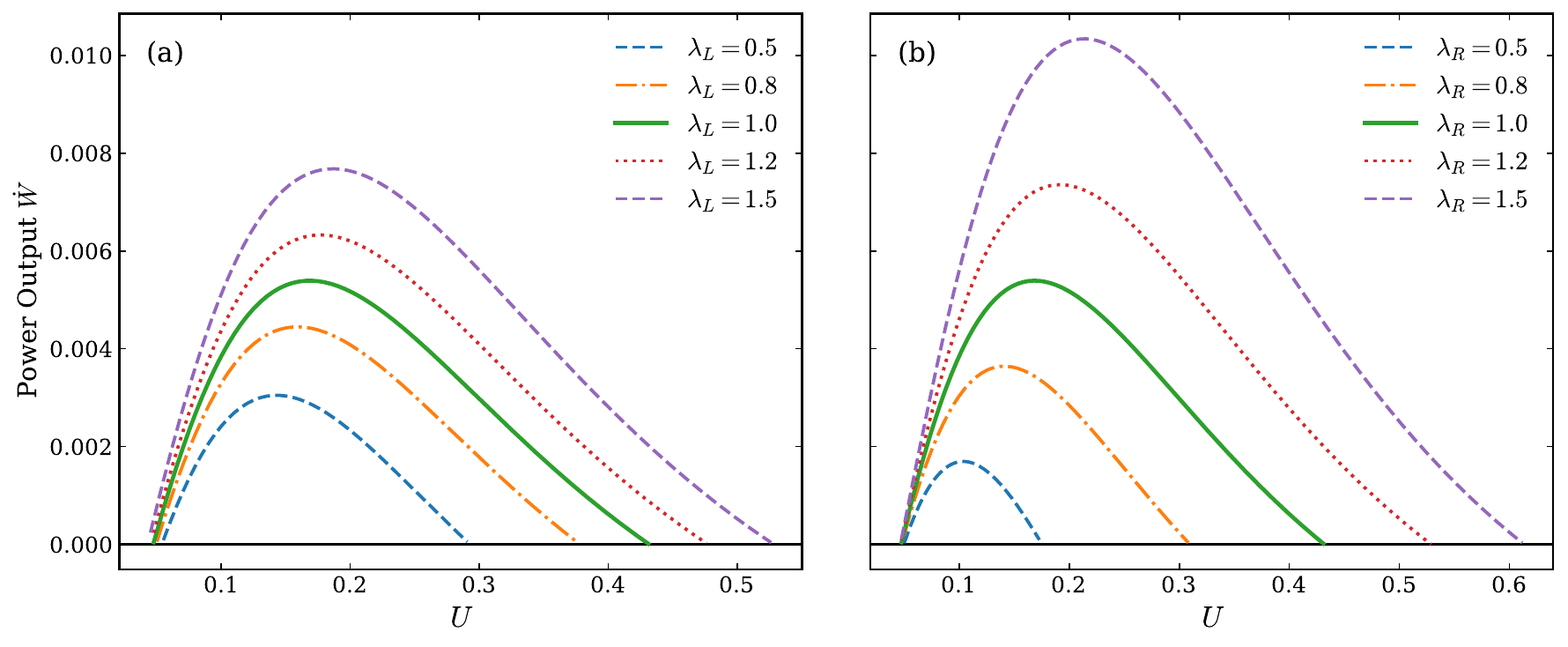}
    \caption{
     Power output $\dot{W}$ under modulation of the renormalized electron-electron coupling strengths.
    (a) Varying $U_L'=\lambda_LU$ with $\lambda_R=1$.
    (b) Varying $U_R'=\lambda_RU$ with $\lambda_L=1$.
    In both plots, the reciprocal reference (green solid curve) corresponds to $\lambda_L=\lambda_R=1$. Curves are shown only in the heat-engine operation regime. All other parameters are fixed to their default values given in Table~\ref{tab:numerical_parameters}.
    }
    \label{fig:power_Uprime}
\end{figure*}

\subsection{Power output and information-to-work conversion}
\label{sec:3D}
We finally examine the power output $\dot{W}$ of the nonreciprocal heat engine. The extracted power is computed from the steady-state energy balance introduced in Sec.~\ref{sec:2c}, and is positive only when the system operates in the heat-engine mode. The definition of $\dot{W}$ includes the energetic contribution associated with maintaining the nonreciprocal transition kinetics. In addition to the power itself, we also evaluate the information-to-work conversion efficacy $\xi=\dot{W}/\dot{I}$ introduced in Sec.~\ref{sec:2c}. For the same reason as in the efficiency analysis, we present the power and conversion efficacy using expressions after the $\epsilon$-independent reparametrization. This representation makes the changes in the power window and in the information-to-work conversion transparent when tuning the nonreciprocity.

A set of numerical results for the power output $\dot{W}$ is shown in Fig.~\ref{fig:power_Uprime}. From both plots, we see that increasing either $\lambda_L$ or $\lambda_R$ in the parameter range shown not only enhances the magnitude of the power but also broaden the positive-power region toward larger $U$. This trend clearly indicates that nonreciprocity can tune the range of electron–electron coupling strengths over which work can be extracted.

\begin{figure*}[t]
    \centering
    \includegraphics[width=0.86\textwidth]{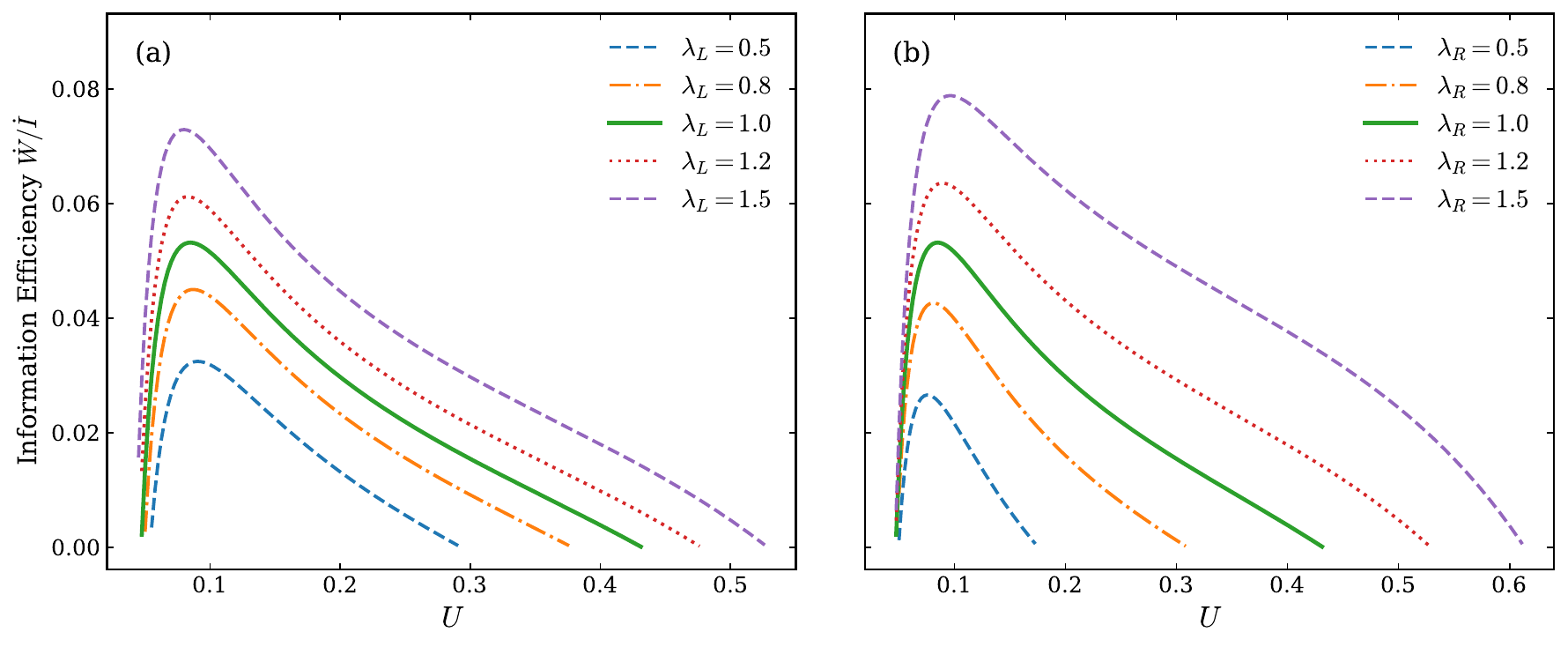}
    \caption{
    Information-to-work conversion efficacy $\xi=\dot{W}/\dot{I}$ under modulation of the renormalized electron-electron coupling strengths.
    (a) Varying $U_L'=\lambda_LU$ with $\lambda_R=1$.
    (b) Varying $U_R'=\lambda_RU$ with $\lambda_L=1$.
    In both plots, the reciprocal reference (green solid curve) corresponds to $\lambda_L=\lambda_R=1$. Curves are shown only in the heat-engine operation regime. All other parameters are fixed to their default values given in Table~\ref{tab:numerical_parameters}.
    }
    \label{fig:WoverI_Uprime}
\end{figure*}

We present numerical results for the information-to-work conversion efficacy in Fig.~\ref{fig:WoverI_Uprime}. We observe that increasing either $\lambda_L$ or $\lambda_R$ can increase both the magnitude of $\xi$ and the range of $U$ over which positive information-to-work conversion is sustained. As derived in Eq.~\eqref{eq:xi_bound}, this conversion efficacy is upper bounded by $T_Y-T_X$, which equals $0.9$ for the parameters used here. This upper bound is not plotted in the figure because the numerical values of $\xi$ are far below it for the chosen parameter sets. Together with Fig.~\ref{fig:power_Uprime}, these results demonstrate that the nonreciprocal heat engine outperforms its reciprocal counterpart when $\lambda_L>1$ or $\lambda_R>1$ in terms of both power output and information-to-work conversion efficacy, consistent with our previous thermodynamic efficiency analysis.

\section{Discussion and conclusion}\label{sec:4}
In this work, we have shown that nonreciprocal dot-reservoir couplings provide an effective mechanism for manipulating information flow and thermodynamic performance in autonomous double-quantum-dot information engines. The steady-state information flow displays a robust single-peak dependence on the electron-electron coupling strength. This behavior reflects a competition between correlation generation and dynamical suppression: weak interactions are insufficient to establish strong conditional correlations between the two dots, whereas strong interactions suppress the global cycle that carries the information flow. Nonreciprocity reshapes this competition by modifying the transition kinetics and therefore changes both the magnitude and the optimal electron-electron coupling strength at which the information flow attains its maximum.

To provide more insights into the impact of nonreciprocity, we have also introduced a reparametrization scheme that maps the nonreciprocal systems onto an effective reciprocal one but with renormalized system and reservoir parameters. The reparameterized description offers a transparent interpretation of the nonreciprocal effect. Part of the effect of nonreciprocity can be absorbed into effective chemical potentials, which is somewhat trivial as one can also shift the chemical potentials in reciprocal systems. The genuinely nonreciprocal feature is instead the decoupling of the effective electron-electron coupling strengths associated with different transition paths. In a reciprocal bipartite engine, all conditional transitions are governed by a single Coulomb interaction energy. By contrast, nonreciprocal couplings allow the effective coupling strengths $U'_X$, $U'_L$, and $U'_R$ to become mutually distinct. This path-dependent splitting enlarges the accessible thermodynamic parameter space and provides a control mechanism that cannot be reproduced by simply tuning the static biases of a reciprocal system.

We have further verified the thermodynamic consistency of the generalized framework. The asymmetric transition rates introduce additional logarithmic contributions to the thermodynamic affinities, which represent the cost required to maintain nonreciprocal kinetics. Including these contributions ensures that the local entropy production rates of both the controller and the working substance remain non-negative. Thus, nonreciprocity modifies the internal current and information-flow dynamics without violating the second law of thermodynamics.

From the perspective of engine performance, nonreciprocity can substantially reshape the efficiency, power output, and information-to-work conversion efficacy. Specifically, the heat-engine operation window can be broadened and the peak power can be enhanced by tuning nonreciprocity, providing a route to optimize the performance of heat engine. Nevertheless, the global thermodynamic efficiency remains bounded by the Carnot limit. Nonreciprocity therefore acts as a resource for redistributing currents and optimizing performance within, rather than beyond, the constraints imposed by the thermodynamics.

Several limitations should be noted. The present study is based on a Markovian master equation description, and the nonreciprocal transition rates are introduced phenomenologically. A microscopic realization of such asymmetric tunneling, for example through dissipation engineering~\cite{Metelmann.15.PRX}, remains an important direction for future work. Moreover, we have focused mainly on the heat-engine regime. Other operation modes, such as refrigerator, or autonomous pumping, may reveal additional roles of nonreciprocal coupling. We leave them for future works.


\begin{acknowledgments}
This work is supported by the National Natural Science Foundation of China (Grant No. 12205179), the Shanghai Pujiang Program (Grant No. 22PJ1403900) and the Shanghai Science and Technology Innovation Action Plan (Grant No. 24LZ1400800).
\end{acknowledgments}

\appendix

\section{Derivation of the entropy production rates in the nonreciprocal system}
\label{app:entropy_derivation}
In this appendix, we provide the detailed derivation of the generalized total and local entropy production rates [Eqs. \eqref{eq:Si17}-\eqref{eq:Siy19}] for the autonomous double-quantum-dot information engine in the presence of nonreciprocal dot-reservoir couplings.

Based on the network theory of stochastic thermodynamics, the total steady-state entropy production rate $\dot{S}_i$ is entirely given by the environmental entropy production $\dot{S}_r$ since the system's Shannon entropy remains constant in the steady state. The total entropy production rate can be expressed as a sum over the three fundamental cycles $\mathcal{C}_n\in\{\mathcal{C},\mathcal{C}_Y^0,\mathcal{C}_Y^1\}$ of the system
\begin{equation}
    \dot{S}_i = \sum_{n} \mathcal{J}(\mathcal{C}_n) \mathcal{F}(\mathcal{C}_n),
\end{equation}
where $\mathcal{J}(\mathcal{C}_n)$ is the steady-state probability current along the cycle $\mathcal{C}_n$, and $\mathcal{F}(\mathcal{C}_n)$ is the corresponding thermodynamic affinity defined by the logarithmic ratio of the product of forward and backward transition rates along the cycle~(see, e.g, Ref.~\cite{PhysRevX.4.031015}).

\subsection*{1. Thermodynamic affinity for local cycles $\mathcal{C}_Y^{0/1}$}
We first consider the local cycle $\mathcal{C}_Y^{x}$, which corresponds to an electron entering the working substance ($Y$ dot) from the right reservoir and subsequently leaving to the left reservoir, while the controller ($X$ dot) remains in a fixed state $x \in \{0, 1\}$. The transition sequence is $(x, 0) \xrightarrow{R} (x, 1) \xrightarrow{L} (x, 0)$. 
Using the generalized nonreciprocal transition rates defined in Eq. (\ref{eq:wx_non}), the thermodynamic affinity for this local cycle reads
\begin{eqnarray}
    \mathcal{F}(\mathcal{C}_Y^x) &\equiv& \ln \left(\frac{W_x^{10,(R)} W_x^{01,(L)}}{W_x^{01,(R)} W_x^{10,(L)}}\right)\nonumber\\
    &=& \ln \left[ \frac{\Gamma_{x+}^{(R)} f_x^{(R)}}{\Gamma_{x-}^{(R)} (1-f_x^{(R)})} \frac{\Gamma_{x-}^{(L)} (1-f_x^{(L)})}{\Gamma_{x+}^{(L)} f_x^{(L)}} \right].
\end{eqnarray}
Substituting the nonreciprocity ratios $\alpha_x^{(\nu)} = \Gamma_{x+}^{(\nu)} / \Gamma_{x-}^{(\nu)}$ and the conditional Fermi-Dirac distributions $f_x^{(\nu)} = \{1 + \exp[(\epsilon_Y + xU - \mu_\nu)/T_Y]\}^{-1}$, we can separate the kinetic and thermodynamic contributions:
\begin{eqnarray}
    \mathcal{F}(\mathcal{C}_Y^x) &=& \ln \frac{\alpha_x^{(R)}}{\alpha_x^{(L)}} + \ln \left[ \exp\left(-\frac{\epsilon_Y + xU - \mu_R}{T_Y}\right) \right.\nonumber\\
    &&\times\left.\exp\left(\frac{\epsilon_Y + xU - \mu_L}{T_Y}\right) \right] \nonumber\\
    &=& \ln \frac{\alpha_x^{(R)}}{\alpha_x^{(L)}} - \frac{\mu_L - \mu_R}{T_Y} \nonumber\\
    &=& \ln \frac{\alpha_x^{(R)}}{\alpha_x^{(L)}} - \frac{\Delta\mu}{T_Y}.
\end{eqnarray}

\subsection*{2. Thermodynamic affinity for the global cycle $\mathcal{C}$}
Next, we evaluate the thermodynamic affinity for the global cycle $\mathcal{C}$, which follows the path $(0,0) \to (1,0) \to (1,1) \to (0,1) \to (0,0)$. This cycle involves transitions of both dots and the exchange of energy and information. The definition for the associated thermodynamic affinity is given by
\begin{equation}
    \mathcal{F}(\mathcal{C}) \equiv \ln \left(\frac{W_{10}^0 W_1^{10,(L)} W_{01}^1 W_0^{01,(L)}}{W_{01}^0 W_1^{01,(L)} W_{10}^1 W_0^{10,(L)}}\right).
\end{equation}
Applying the nonreciprocal transition rates in Eqs. (\ref{eq:wy_non}) and (\ref{eq:wx_non}), we get
\begin{eqnarray}
    \mathcal{F}(\mathcal{C}) &=& \ln \left[ \left( \frac{\Gamma^{0+}}{\Gamma^{0-}} \frac{\Gamma_{1+}^{(L)}}{\Gamma_{1-}^{(L)}} \frac{\Gamma^{1-}}{\Gamma^{1+}} \frac{\Gamma_{0-}^{(L)}}{\Gamma_{0+}^{(L)}} \right)\right.\nonumber\\ 
    &&\times\left.\left( \frac{f_0}{1-f_0} \frac{f_1^{(L)}}{1-f_1^{(L)}} \frac{1-f_1}{f_1} \frac{1-f_0^{(L)}}{f_0^{(L)}} \right) \right].
\end{eqnarray}
The term in the first bracket simplifies to $\frac{\gamma^0 \alpha_1^{(L)}}{\gamma^1 \alpha_0^{(L)}}$. Evaluating the Fermi-Dirac distribution ratios yields the thermal bias introduced by the electron-electron coupling
\begin{eqnarray}
    \ln \left( \frac{f_0}{1-f_0} \frac{1-f_1}{f_1} \right) &=& -\frac{\epsilon_X - \mu_X}{T_X} + \frac{\epsilon_X + U - \mu_X}{T_X} \nonumber\\
    &=& \frac{U}{T_X},\nonumber \\
    \ln \left( \frac{f_1^{(L)}}{1-f_1^{(L)}} \frac{1-f_0^{(L)}}{f_0^{(L)}} \right) &=& -\frac{\epsilon_Y + U - \mu_L}{T_Y} + \frac{\epsilon_Y - \mu_L}{T_Y} \nonumber\\
    &=& -\frac{U}{T_Y}.
\end{eqnarray}
Consequently, the associated thermodynamic affinity of the global cycle finally reads
\begin{equation}
    \mathcal{F}(\mathcal{C}) = \ln \frac{\gamma^0 \alpha_1^{(L)}}{\gamma^1 \alpha_0^{(L)}} + \frac{U}{T_X} - \frac{U}{T_Y}.
\end{equation}
Combining the current and affinity contributions from the local and global cycles, the total steady-state EPR is obtained
\begin{eqnarray}
    \dot{S}_i &=& \sum_{x=0,1} \mathcal{J}(\mathcal{C}_Y^x) \left( \ln \frac{\alpha_x^{(R)}}{\alpha_x^{(L)}} - \frac{\Delta\mu}{T_Y} \right) \nonumber\\
    &&+ \mathcal{J}(\mathcal{C}) \left( \ln \frac{\gamma^0 \alpha_1^{(L)}}{\gamma^1 \alpha_0^{(L)}} + \frac{U}{T_X} - \frac{U}{T_Y} \right),
\end{eqnarray}
which is just Eq. (\ref{eq:Si17}) in the main text.

\subsection*{3. Evaluating local entropy production rates}
We now turn to local entropy production rates associated with each dot. The local environmental entropy production rate for the controller ($X$ dot) depends solely on the X-only transitions along the global cycle,
\begin{equation}
    \dot{S}_r^X = \mathcal{J}(\mathcal{C}) \ln \frac{W_{10}^0 W_{01}^1}{W_{01}^0 W_{10}^1} = \mathcal{J}(\mathcal{C}) \left( \ln \frac{\gamma^0}{\gamma^1} + \frac{U}{T_X} \right).
\end{equation}
By inserting the continuous information flow $\dot{I} = \mathcal{J}(\mathcal{C})\mathcal{F}^I(\mathcal{C})$ that applies to nonreciprocal systems, the local entropy production rate for $X$ dot still satisfies the generalized second law locally, leading to Eq. (\ref{eq:Six18}) in the main text,
\begin{equation}
    \dot{S}_i^X = \dot{S}_r^X - \dot{I} = \mathcal{J}(\mathcal{C}) \left[ \ln \frac{\gamma^0}{\gamma^1} + \frac{U}{T_X} - \mathcal{F}^I(\mathcal{C}) \right].
\end{equation}

Similarly, the environmental entropy production rate for the working substance ($Y$ dot) includes contributions from both the local cycles and the Y-only transitions within the global cycle
\begin{equation}
    \dot{S}_r^Y = \sum_x \mathcal{J}(\mathcal{C}_Y^x) \mathcal{F}(\mathcal{C}_Y^x) + \mathcal{J}(\mathcal{C}) \ln \frac{W_1^{10,(L)} W_0^{01,(L)}}{W_1^{01,(L)} W_0^{10,(L)}}.
\end{equation}
Evaluating the contribution from the global cycle for $Y$ dot yields
\begin{eqnarray}
    \ln \frac{W_1^{10,(L)} W_0^{01,(L)}}{W_1^{01,(L)} W_0^{10,(L)}} &=& \ln \left[ \frac{\alpha_1^{(L)}}{\alpha_0^{(L)}} \exp\left(-\frac{\epsilon_Y + U - \mu_L}{T_Y}\right) \right.\nonumber\\
    &&\times\left.\exp\left(\frac{\epsilon_Y - \mu_L}{T_Y}\right) \right] \nonumber\\
    &=& \ln \frac{\alpha_1^{(L)}}{\alpha_0^{(L)}} - \frac{U}{T_Y}.
\end{eqnarray}
Taking the information flow into account, we arrive at the expression for the local entropy production rate for $Y$ dot, $\dot{S}_i^Y = \dot{S}_r^Y + \dot{I}$, as presented in Eq. (\ref{eq:Siy19}) in the main text
\begin{eqnarray}
    \dot{S}_i^Y &=& \sum_x \mathcal{J}(\mathcal{C}_Y^x) \left[ \ln \frac{\alpha_x^{(R)}}{\alpha_x^{(L)}} - \frac{\Delta\mu}{T_Y} \right] \nonumber\\
    &&+ \mathcal{J}(\mathcal{C}) \left[ \ln \frac{\alpha_1^{(L)}}{\alpha_0^{(L)}} - \frac{U}{T_Y} + \mathcal{F}^I(\mathcal{C}) \right].
\end{eqnarray}
This completes the derivation, showing how logarithms of the nonreciprocal coupling act as actively sustained thermodynamic forces that restructure the entropy balance.

\section{$\epsilon$-dependent reparameterization scheme}
\label{app:leadindep}
This appendix provides details of the $\epsilon$-dependent reparameterization scheme, which complements the $\epsilon$-independent scheme presented in the main text. For the controller ($X$ dot), we introduce the renormalized effective energy level $\epsilon_X'$ and the effective electron-electron coupling strength $U_X'$ as
\begin{equation}
    \epsilon_X' = \epsilon_X - T_X \ln \gamma^0, \quad U_X' = U - T_X \ln \frac{\gamma^1}{\gamma^0}.
\end{equation}
Compared with bare parameters of the original reciprocal system, the renormalized parameters become proportional to the temperature $T_X$ of the reservoir coupled to $X$ dot, with prefactors that depend on the nonreciprocity ratios $\gamma^{0/1}$. The corresponding effective Fermi-Dirac distribution is given by $f_y' = [1 + \exp((\epsilon_X' + yU_X' - \mu_X)/T_X)]^{-1}$. With the above renormalized parameters, we find that the generalized transition rates for $X$ dot in Eq. (\ref{eq:wy_non}) can be expressed as
\begin{equation}
    W_{10}^y = \Gamma^{y'} f_y', \quad W_{01}^y = \Gamma^{y'} (1 - f_y').
\end{equation}
Here, the effective reciprocal tunneling rate $\Gamma^{y'} = \Gamma^{y-} + (\Gamma^{y+} - \Gamma^{y-})f_y$ remains the same as in the $\epsilon$-independent scheme presented in the main text. 

For the working substance ($Y$ dot), we introduce the following set of renormalized parameters
\begin{eqnarray}
    \epsilon_Y' &=& \epsilon_Y - T_Y \ln \alpha_0^{(L)}, \quad U_Y' = U - T_Y \ln \frac{\alpha_1^{(L)}}{\alpha_0^{(L)}},\nonumber\\
    \mu_L' &=& \mu_L, \quad \mu_R' = \mu_R + T_Y \ln \frac{\alpha_0^{(R)}}{\alpha_0^{(L)}}.
\end{eqnarray}
Similar to $X$ dot, the renormalized parameters (except $\mu_L'$) become temperature-dependent as well, with nonreciprocity ratios determining the prefactors. In this scheme, the renormalized distribution $f_x'$ takes a piecewise form depending on the state of $X$ dot and the specific reservoir involved
\begin{equation}
    f_x' = \begin{cases}
        \left[ 1 + \exp \left( \frac{\epsilon_Y' + xU_Y' - \mu_R'}{T_Y} \right) \right]^{-1}, & x=0 \text{ and } \nu=R \\
        \left[ 1 + \exp \left( \frac{\epsilon_Y' + xU_Y' - \mu_L}{T_Y} \right) \right]^{-1}, & \text{otherwise.}
    \end{cases}
\end{equation}
With the above renormalized parameters, the generalized transition rates for $Y$ dot in Eq. (\ref{eq:wy_non}) are then expressed as $W_x^{10,\nu} = \Gamma_x^{(\nu)'} f_x'$ and $W_x^{01,\nu} = \Gamma_x^{(\nu)'} (1 - f_x')$, where $\Gamma_x^{(\nu)'}$ is the same effective reciprocal tunneling rate defined in the $\epsilon$-independent scheme presented in the main text.

Within this $\epsilon$-dependent scheme, the expressions for the steady-state local EPRs for the two dots in Eqs.~(\ref{eq:Six18}) and (\ref{eq:Siy19}) can be recast as
\begin{eqnarray}
    \dot{S}_i^X &=& \mathcal{J}(C) \left[ \frac{U_X'}{T_X} - \mathcal{F}^I(C) \right], \\
    \dot{S}_i^Y &=& -\mathcal{J}(C_Y^0) \frac{\Delta \mu}{T_Y} - \mathcal{J}(C_Y^1) \left( \frac{\Delta \mu}{T_Y} - \xi \right) \nonumber\\
    &&+ \mathcal{J}(C) \left[ \mathcal{F}^I(C) - \frac{U_Y'}{T_Y} \right],
\end{eqnarray}
where $\Delta \mu = \mu_L - \mu_R'$ represents the renormalized chemical potential bias. We have also denoted $\xi$ as
\begin{equation}
    \xi = \ln \frac{\alpha_1^{(R)} \alpha_0^{(L)}}{\alpha_0^{(R)} \alpha_1^{(L)}},
\end{equation}
which depends only on nonreciprocity ratios and vanishes in the reciprocal limit. These expressions highlight that while $U_X'$ and $U_Y'$ act as effective interaction energies, $\xi$ functions as an additional tunable thermodynamic force that modifies the entropy balance of the engine. Compared with Eqs. (\ref{eq:sxrepar}) and (\ref{eq:syrepar}) of the $\epsilon$-independent scheme, we find that the expression for $dot{S}_i^Y$ becomes more complicated.


%

\end{document}